\documentclass[sigconf]{acmart} %authordraft,
\UseRawInputEncoding
\usepackage{amsthm}
\usepackage{amsfonts} % amssymb
\usepackage{url}          
\usepackage{ifthen}
\usepackage{color, xcolor}
\hypersetup{colorlinks=true, linkcolor=blue, citecolor=blue, anchorcolor=blue, urlcolor=blue}% link color
\usepackage{thm-restate}
\usepackage{algorithm}
\usepackage{algorithmic}
\usepackage{enumitem}
\usepackage{multirow}
\usepackage{float}
\usepackage{stfloats}
\usepackage{subfig}

\renewcommand\footnotetextcopyrightpermission[1]{}  % remove copyright
\AtBeginDocument{%
  \providecommand\BibTeX{{%
    \normalfont B\kern-0.5em{\scshape i\kern-0.25em b}\kern-0.8em\TeX}}}

\setcopyright{acmcopyright}
\copyrightyear{2022}
\acmYear{2022}
\acmDOI{XXXXXXX.XXXXXXX}

\acmConference[ACL '23]{The 61st Annual Meeting of the Association for Computational Linguistics}{July 9--14, 2023}{Toronto, Ontario, Canada}
\acmPrice{15.00}
\acmISBN{978-1-4503-XXXX-X/18/06}

\newcommand{\compilehidecomments}{false}
\ifthenelse{ \equal{\compilehidecomments}{true} }{%
	\newcommand{\wei}[1]{}
	\newcommand{\sheng}[1]{}
	\newcommand{\hao}[1]{}
	\newcommand{\zhou}[1]{}
        \newcommand{\zhang}[1]{}
}{
	\newcommand{\wei}[1]{{\color{blue!50!black}  [\text{Wei:} #1]}}
	\newcommand{\sheng}[1]{{\color{red!70!black} [\text{Sheng:} #1]}}
	\newcommand{\hao}[1]{{\color{green!90!black} [\text{Hao:} #1]}}
	\newcommand{\zhou}[1]{{\color{yellow!90!black} [\text{zhou:} #1]}}
        \newcommand{\zhang}[1]{{\color{purple!90!black} [\text{zhang:} #1]}}
}

\begin{document}

\title{Explainable Recommendation with Personalized Review Retrieval and Aspect Learning}

\author{Hao Cheng}
\affiliation{%
  \institution{Shenzhen University}
  \city{Shenzhen}
  \country{China}}
  \email{2110276103@email.szu.edu.cn}

\author{Shuo Wang}
\affiliation{%
  \institution{Shenzhen University}
  \city{Shenzhen}
  \country{China}}
\email{2110276109@email.szu.edu.cn}

\author{Wensheng Lu}
\affiliation{%
  \institution{Shenzhen University}
  \city{Shenzhen}
  \country{China}}
\email{2210273060@email.szu.edu.cn}

\author{Wei Zhang}
\affiliation{%
  \institution{Shenzhen University}
  \city{Shenzhen}
  \country{China}}
\email{2210275010@email.szu.edu.cn}

\author{Mingyang Zhou}
\affiliation{%
  \institution{Shenzhen University}
  \city{Shenzhen}
  \country{China}}
\email{zmy@szu.edu.cn}

\author{Kezhong Lu}
\affiliation{%
  \institution{Shenzhen University}
  \city{Shenzhen}
  \country{China}}
\email{kzlu@szu.edu.cn}

\author{Hao Liao}
\authornote{Corresponding author}
\affiliation{%
  \institution{Shenzhen University}
  \city{Shenzhen}
  \country{China}}
\email{haoliao@szu.edu.cn}

\renewcommand{\shortauthors}{Hao Cheng et al.}

\begin{abstract}
Explainable recommendation is a technique that combines prediction and generation tasks to produce more persuasive results. Among these tasks, textual generation demands large amounts of data to achieve satisfactory accuracy. However, historical user reviews of items are often insufficient, making it challenging to ensure the precision of generated explanation text. To address this issue, we propose a novel model, ERRA (\textbf{E}xplainable \textbf{R}ecommendation by personalized \textbf{R}eview retrieval and \textbf{A}spect learning). With retrieval enhancement, ERRA can obtain additional information from the training sets. With this additional information, we can generate more accurate and informative explanations. Furthermore, to better capture users' preferences, we incorporate an aspect enhancement component into our model. By selecting the top-$n$ aspects that users are most concerned about for different items, we can model user representation with more relevant details, making the explanation more persuasive.  To verify the effectiveness of our model, extensive experiments on three datasets show that our model outperforms state-of-the-art baselines (for example, 3.4\% improvement in prediction and 15.8\% improvement in explanation for TripAdvisor).
\end{abstract}

\maketitle

\section{Introduction}

Recent years have witnessed a growing interest in the development of explainable recommendation models \cite{ref1,ref2}. In general, there are three different kinds of frameworks for explainable recommendation models, which are post-hoc \cite{ref3}, embedded \cite{ref4}, and multi-task learning methods\cite{ref5}. Post-hoc methods generate explanations for a pre-trained model after the fact, leading to limited diversity in explanations. Embedded methods, on the other hand, demonstrate efficacy in acquiring general features from samples and mapping data to a high-dimensional vector space. However, since embedded methods rely on historical interactions or features to learn representations, they may struggle to provide accurate recommendations for users or items with insufficient data.

In addition to the two frameworks mentioned above, there has been a utilization of multi-task learning frameworks in explainable recommendation systems, where the latent representation shared between user and item embeddings is employed \cite{ref5, ref1}. These frameworks often employ the Transformer \cite{ref8,ref9}, a powerful text encoder and decoder structure widely used for textual processing tasks. While efficient for prediction tasks, they encounter challenges in generation tasks due to limited review content, leading to a significant decline in performance. Furthermore, these previous transformer-based frameworks do not incorporate personalized information and treat heterogeneous textual data indiscriminately. To address these issues, we make adaptations to the existing multi-task learning framework by incorporating two main components: retrieval enhancement, which alleviates the problem of data scarcity, and aspect enhancement, which facilitates the generation of specific and relevant explanations.

\begin{figure}[t]
\centering
\includegraphics[height=5cm,width=8cm]{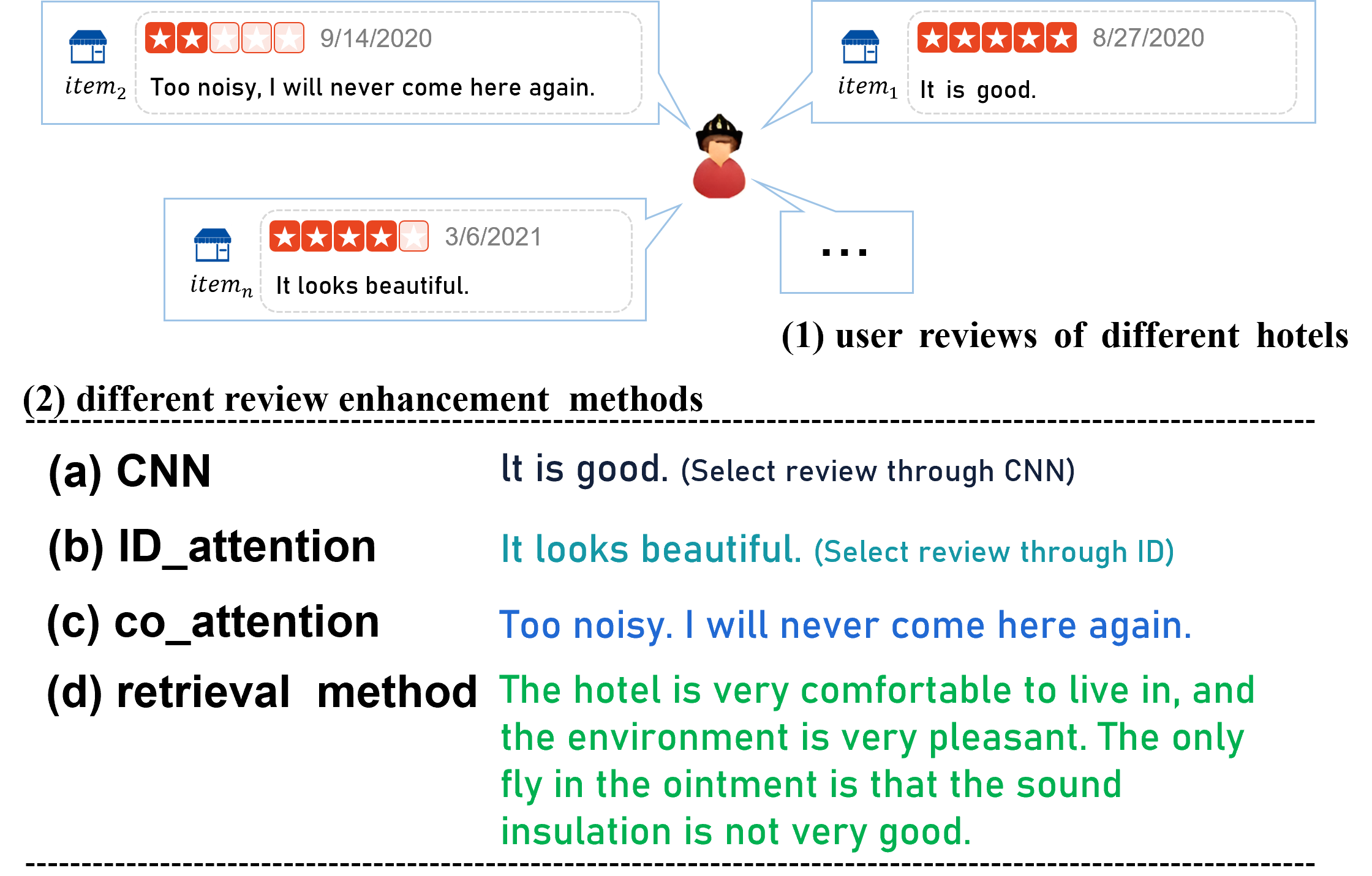}
\vspace{-0.2cm}
\caption{A user's reviews of different items and selected reviews by different models. Specifically, (a) a CNN-based method, by which the review selected is too general, (b) a user-id attention-based query method \cite{ref29}, by which the review selected is not specific, (c) a Co-attention based method \cite{ref5}, by which the review selected contain some details, (d) our model: retrieval-based method generates informative and personalized reviews that are relevant to the hotel.}
\label{fig:motivation}
\end{figure}

Real-world datasets usually contain redundant reviews generated by similar users, making the selected reviews uninformative and meaningless, which is illustrated in Figure \ref{fig:motivation}. To address this issue, a model-agnostic retrieval enhancement method has been employed to identify and select the most relevant reviews. Retrieval is typically implemented using established techniques, such as TF-IDF (Term Frequency-Inverse Document Frequency) or BM25 (Best Match 25) \cite{ref6}, which efficiently match keywords with an inverted index and represent the question and context using high-dimensional sparse vectors. This approach facilitates the generation of sufficient specific text, thereby attaining enhanced textual quality for the user. Generally, Wikipedia is utilized as a retrieval corpus for the purpose of aiding statement verification \cite{ref33,ref34}. Here, we adopt a novel approach wherein the training set of each dataset is utilized as the retrieval corpus.  By integrating this component into our framework, we are able to generate sentences with more specific and relevant details. Consequently, this enhancement facilitates the generation of explanations that are more accurate, comprehensive, and informative at a finer granularity.

Moreover, users rarely share a common preference \cite{ref29}. Therefore, aspects \cite{ref12}, extracted from corresponding reviews, can be utilized to assist in the modeling of user representation. The incorporation of aspect enhancement has resulted in not only improved prediction accuracy but also more personalized and user-specific text during the text generation process.

By incorporating retrieval enhancement and aspect enhancement into our model, we adjust the transformer architecture to meet our needs, achieving better performance in both prediction and generation tasks.

The main contributions of our framework are as follows:
\begin{itemize}[leftmargin=*]

\item In response to the problem of insufficient historical reviews for users and items in explainable recommendation systems, we propose a retrieval enhancement technique to supplement the available information with knowledge bases obtained from a corpus. To the best of our knowledge, this study represents the first application of retrieval-enhanced techniques to review-based explainable recommendations. 
\item We propose a novel approach wherein different aspects are selected for individual users when interacting with different items and are subsequently utilized to facilitate the modeling of user representation, thereby leading to the generation of more personalized explanations.

\item Experimental results on real-world datasets demonstrate the effectiveness of our proposed approach, achieving superior performance compared to state-of-the-art baselines\footnote{https://github.com/lileipisces/PETER}.
\end{itemize}

% \vspace{-4mm}
\section{Related Work}

\subsection{Explainable Recommendation with Generation}

Explainable recommendation systems \cite{ref38} have been extensively studied using two primary methodologies: machine learning and human-computer interaction. The former \cite{ref39, ref40} investigates how humans perceive different styles of explanations, whereas the latter generates explanations through the application of explainable recommendation algorithms, which is more relevant to our research. Numerous approaches exist for explaining recommendations, including the use of definition templates \cite{ref41}, image visualization \cite{ref42}, knowledge graphs \cite{ref43}, and rule justifications \cite{ref44}. Among these methods, natural language explanations \cite{ref5, ref9} are gaining popularity due to their user accessibility, advancements in natural language processing techniques, and the availability of vast amounts of text data on recommendation platforms. Several studies have employed Recurrent Neural Network (RNN) networks \cite{ref23}, coupled with Long Short-Term Memory (LSTM) \cite{ref48}, for generating explanatory texts, while others have utilized co-attention and Gated Recurrent Unit (GRU) \cite{ref49} in conjunction with Convolutional Attentional Memory Networks (CAML) \cite{ref5} for text generation. More recently, transformer-based networks have seen increased utilization for score prediction and interpretation generation. \cite{ref9}

\textbf{Pre-trained Models}
The pre-trained model has gained significant traction in the field of NLP recently. These models, such as \cite{ref32,ref11} are trained on large-scale open-domain datasets utilizing self-supervised learning tasks, which enables them to encode common language knowledge. The ability to fine-tune these models with a small amount of labeled data has further increased their utility for NLP tasks \cite{ref14,ref17}. For example, a pre-trained model is Sentence-BERT \cite{ref11}, which utilizes a multi-layer bidirectional transformer encoder and incorporates the Masked Language Model and Next Sentence Prediction to capture word and sentence-level representations. Another example is UniLM \cite{ref13}, which builds upon the architecture of BERT and has achieved outstanding performance in a variety of NLP tasks including unidirectional, bidirectional, and sequence-to-sequence prediction. Furthermore, research has demonstrated that pre-trained models possess the capability to capture hierarchy-sensitive and syntactic dependencies \cite{ref14}, which is highly beneficial for downstream NLP tasks. The utilization of pre-trained models has proven to be a powerful approach in the NLP field, with the potential to further improve performance on a wide range of tasks.

\noindent\textbf{Retrieval Enhancement} 
Retrieval-enhanced text generation has recently received increased attention due to its capacity to enhance model performance in a variety of natural language processing (NLP) tasks ~\cite{ref17,ref14}. For instance, in open-domain question answering, retrieval-enhanced text generation models can generate the most up-to-date answers by incorporating the latest information during the generation process~\cite{ref15,ref16}. This is not possible for traditional text generation models, which store knowledge through large parameters, and the stored information is immutable. Retrieval-based methods also have an advantage in scalability, as they require fewer additional parameters compared to traditional text generation models~\cite{ref17}. Moreover, by utilizing relevant information retrieved from external sources as the initial generation condition~\cite{ref17}, retrieval-enhanced text generation can generate more diverse and accurate text compared to text generation without any external information.

% \vspace{-2mm}
\section{Problem Statement}
\label{sec:Problem statement}
Our task is to develop a model that can accurately predict ratings for a specific product and provide a reasonable explanation for the corresponding prediction. 

The model's input is composed of various elements, namely the user ID, item ID, aspects, reviews, and retrieval sentences, whereas the resulting output of the model encompasses both a prediction and its explanation. We offer a detailed description of our models' input and output data in this section.

\noindent\textbf{Input Data}
\begin{itemize}[leftmargin=*]
\item \textbf{Heterogeneous information}: 

The variables included in the framework encompass user ID $u$, item ID $v$, aspects $A$, retrieval sentences $S$, and review $R$. Aspects $A$ are captured in the form of a vector representing the user's attention, denoted as $(A_{u,1},\ldots, A_{u,n})$, where $A_{u,j}$ represents the $j$-th aspect extracted from the reviews provided by user $u$. As an illustration, the review \emph{The screen of this phone is too small} encompasses the aspect \emph{(screen, small)}. Regarding users, we extract the most important sentence $S_{u,j}$ from the set $(S_{u,1},...,S_{u,n})$. Similar operations are performed for items, where $S_{v,j}$ is employed. Ultimately, the user's review for the item ${R}_{u, v}$ is fed into the training process to enhance the ability to generate sentences. 
\end{itemize}

\noindent\textbf{Output Data}
\begin{itemize}[leftmargin=*]
\item \textbf{Prediction and explaination}: Given a user $u$ and an item $v$, we can obtain a rating prediction $\hat{r}_{u, v}$, representing user $u$'s preference towards item $v$ and a generated explanatory text $\mathbf{L}=(l_1,l_2,\ldots,l_T)$, providing a rationale for the prediction outcome. In this context, $l_i$ denotes the $i$-th word within the explanation text, while $T$ represents the maximum length of the generated text.
\end{itemize}

\section{Methodology}

\subsection{Overview of Model}
Here we present a brief overview of the ERRA model. As shown in Figure \ref{fig:model}, our model mainly consists of three components, each corresponding to a sub-process of the information processing model:
\begin{itemize}[leftmargin=*]
\item \textbf{Retrieval Enhancement} aims to retrieve external knowledge from the training sets.
\item \textbf{Aspect Enhancement} aims to identify the most important aspects that users are concerned about in their reviews. 
\item \textbf{Joint Enhancement Transformers} is responsible for the integration of the retrieved sentences and aspects with a transformer structure for simultaneously performing the prediction and explanation tasks.
\end{itemize}
Next, we will provide an in-depth description of each component and how they are integrated into a unified framework.
\subsection{Retrieval Enhancement}

A major challenge in generating rich and accurate explanations for users is the lack of sufficient review data. However, this problem can be alleviated via retrieval-enhanced technology, which introduces external semantic information. 

\subsubsection{Retrieval Encode}
The retrieval corpus is constructed using the training set. To obtain fine-grained information, lengthy reviews are divided into individual sentences with varied semantics. Using these sentences as the search unit allows the model to generate more fine-grained text. Sentence-BERT \cite{ref11} is utilized to encode each sentence in the corpus, which introduces no additional model parameters. We did not use other LLMs (Large Language Models) for retrieval encoding because it is optimized for dense retrieval and efficient for extensive experiments. Sentence-BERT is considerably faster than BERT or RoBERTa when encoding large-scale sentences and possesses an enhanced capacity for capturing semantic meaning, making it particularly well-suited for the retrieval task. The encoded corpus is saved as an embedding file, denoted as $C$. During the retrieval process, the most relevant information is directly searched from the saved vector $C$, which greatly improves the efficiency of retrieval. 

\subsubsection{Retrieval Method}

We adopt a searching model commonly used in the field of question answering (QA) and utilize cosine similarity for querying as a simple and efficient retrieval method. Here, we use the average of the review embedding $U_{avg}$ of each user as the query. This representation is in the same semantic space and also captures user preferential information to a certain extent. The average embedding $U_{avg}$ of all the reviews for a user is used as a query to retrieve the most similar $n$ sentences $(S_{u,1},..., S_{u,n})$ in the previous corpus $C$.  Our approach incorporates the Approximate Nearest Neighbor (ANN) search technique, with an instantiation based on the Faiss\footnote{https://github.com/facebookresearch/faiss} library to improve retrieval speed through index creation. This optimization substantially decreases the total retrieval search duration. Then, in our implementation, we set $n$ as $3$ and stitch these sentences together to form a final sentence. Sentence-BERT is then used to encode this final sentence to obtain a vector $S_{u,v}$, which represents the user for the item retrieval. Similarly, $S_{v,u}$ is used for items to retrieve users.

\subsection{Aspect Enhancement}
Users' preferences are often reflected in their reviews. To better represent users, we need to select the most important aspects of their reviews. Specifically, we first extract aspects from each user and item review using extraction tools. The extracted aspects from user reviews represent the style of the users in their reviews, while the extracted aspects from item reviews represent the most important features of the item. 
We aim to identify the most important aspects that users are concerned about in their reviews. It is worth noting that users' interests may vary in different situations. For example, when choosing a hotel, a user may care more about the environment. Whereas, price is a key factor to consider when buying a mobile phone. 
To address this, we use the average vector $A_{v_{i},avg}, v_{i} \in V$, representing all aspects under the item reviews, as the query. This vector is encoded using Sentence-BERT. For each user, we construct a local corpus of their aspects collection $(A_{u_{i},1},..., A_{u_{i},l}), u_{i} \in U$ and use cosine similarity as the measurement indicator. We search for the top-n aspects from the local corpus by $A_{v_{i},avg}$. These retrieved aspects represent the top-n aspects that the user is concerned about this item.

\begin{figure}[t]
\setlength{\textfloatsep}{0pt}
\centering
\includegraphics[width=0.50\textwidth]{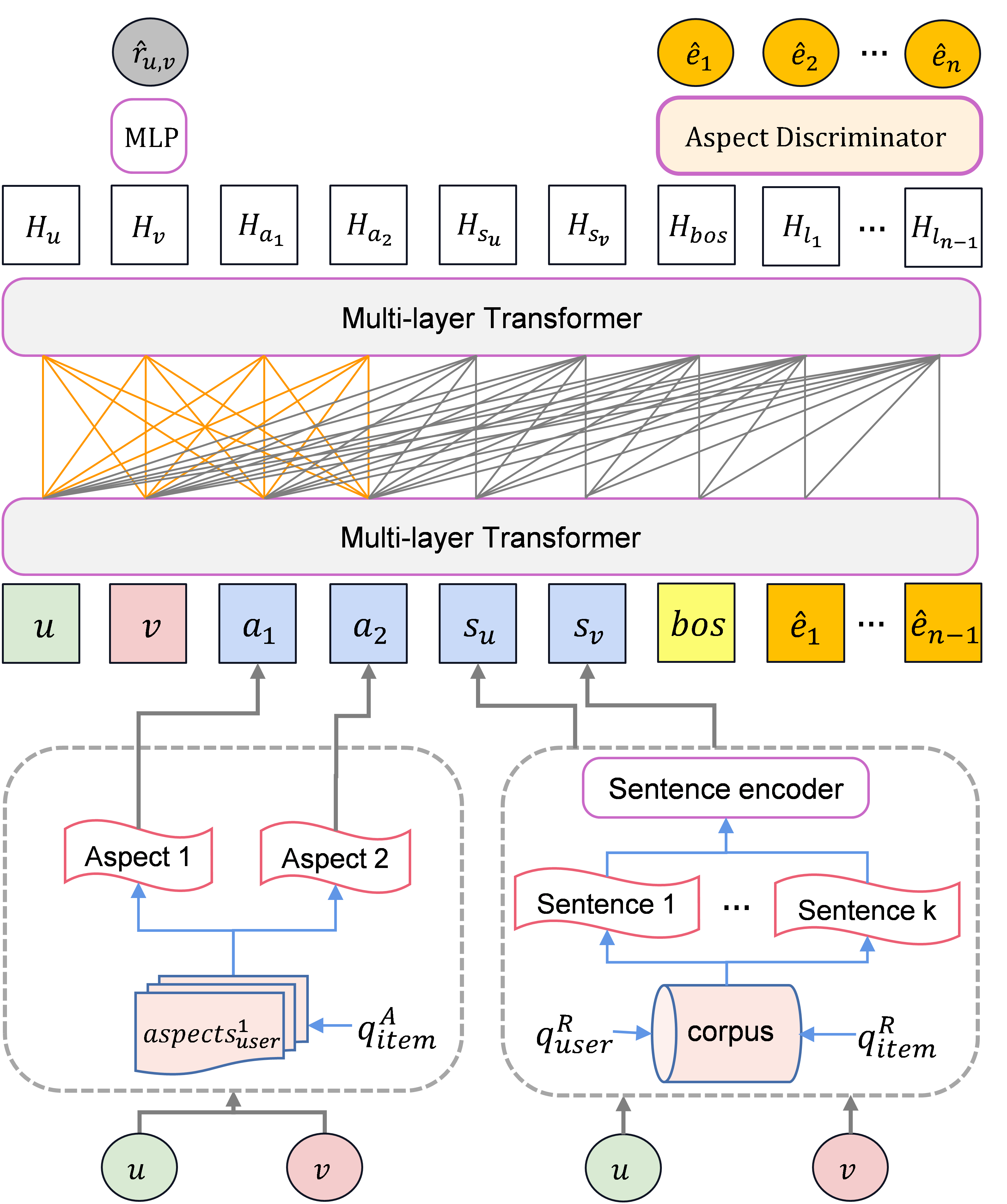}
\setlength{\textfloatsep}{5pt}
\vspace{-0.6cm}
\caption{An overview of the ERRA framework.}
\vspace{-0.6cm}
\label{fig:model}
\label{framework}
\end{figure}

\subsection{Joint Enhancement Transformers}

In our proposed model, we adapt the transformer structure to the prediction and explanation tasks. The transformer consists of multiple identical layers with each layer comprising two sub-layers: the multi-head self-attention and the position-wise feed feedback network. Previous research has made various modifications to the transformer architecture \cite{ref9,ref18}.
Here we integrate the retrieved aspects with the encoding of multiple sentences in various ways. The retrieved sentences $S_{U,j}$, $S_{V,j}$ are encoded uniformly as the input hidden vector $s_{u_{v}}$, $s_{v_{u}}$ and are introduced into the first layer of the transformer.

Below, we use one layer as an example to introduce our calculation steps.

\begin{equation}
\mathbf{A}_{i, h} =\operatorname{softmax}\left(\frac{\mathbf{Q}_{i, h} \mathbf{K}_{i, h}^{\top}}{\sqrt{d}}\right) \mathbf{V}_{i, h} 
\end{equation}
\begin{equation}
\mathbf{Q}_{i, h} =\mathbf{S}_{i-1} \mathbf{W}_{i, h}^{Q}, \mathbf{K}_{i, h}=\mathbf{S}_{i-1} \mathbf{W}_{i, h}^{K}, 
\end{equation}
\begin{equation}
\mathbf{V}_{i, h} =\mathbf{S}_{i} \mathbf{W}_{i, h}^{V}
\end{equation}

where $\mathbf{S}_{i-1}\in{\mathbb{R}}^{|S|\times d}$ is the $i$-th layer's output,  $\mathbf{W}_{i, h}^{Q}, \mathbf{W}_{i, h}^{K}, \mathbf{W}_{i, h}^{V}\in{\mathbb{R}}^{ d \times\frac{d}{H}}$ are projection matrices, d denotes the dimension of embeddings and is set to 384. |S| denotes the length of the input sequence.

Subsequently, we incorporate aspect information into the model. As aspects are closely related to both users and items, we modify the internal mask structure of the model and combine the user's aspects and ID information through a self-attention mechanism. Not only does this strategy account for the uniqueness of the ID when modeling users but also increase the personalization of the user's interactions with the item. Specifically, the same user may have different points of attention when interacting with different items. As illustrated in Figure \ref{fig:model}, we make the information of the first four positions attend to each other, because the first two positions encode the unique user identifier, while the third and fourth positions encapsulate the personalized aspects of the user's preferences. The underlying rationale for selecting these positions is to facilitate the attention mechanism in capturing the interactions between users and products, ultimately enhancing the model's accuracy.  At this point, our final input is as follows: $ [U_{id}, V_{id}, A_{u_{1}}$, $A_{u_{2}}, s_{u_{v}},s_{v_{u}}, t_{1},\ldots,t_{|t_{len}|}] $. After including the location $ [P_{1}, P_{2},P_{3},\ldots,P_{|s|}] $, where $|s|$ is the length of the input, the final input becomes $ [H_{1}, H_{2},H_{3},\ldots,H_{|s|}] $.

For the two different information of ID and aspects, we use them jointly to represent the user and item. We use the self-attention mechanism to combine these two different semantic information, however, we found that it causes the final ID embedding matrix to be very close to the word embedding matrix, resulting in the loss of unique ID information and high duplication in generated sentences. To address this problem, we adopt the strategy from previous research \cite{ref18} that only uses an ID to generate texts, and compares the generated text with the real text to compute the loss $\mathcal{L}_{c}$. To a certain extent, this method preserves the unique ID information in the process of combining aspects, thereby reducing the problem of repetitive sentences.

\begin{equation}
\mathcal{L}_{c}=\sum_{(u, v) \in \mathcal{T}} \frac{1}{\left|t_{len}\right|} \sum_{t=1}^{\left|t_{len}\right|}-\log H_{v}^{g_{ti}}
\end{equation}

where $\mathcal{T}$ denotes the training set. $g_{ti}$ denotes that only use the hidden vector of the position $H_{v}$ to generate the i-th word, $i \in {1,2...,t_{len}}$.

\subsection{Rating Prediction}
We utilized the two positions of the final layer (denoted as $H_{v}$) as the input. To combine the information of the ID and the hidden vector $H_{v}$, we employed a multi-layer perceptron (MLP) to map the input into a scalar. The loss function used in this model was the Root Mean Square Error (RMSE) function, which was used to calculate the loss. 

\begin{equation}
\mathrm{r}_{u, v}=\operatorname{ReLU}\left(\mathbf[H_{v},{u}_{id},{v}_{id}] \mathbf{W}_{l, 1}\right) \mathbf{W}_{l, 2}
\end{equation}

\begin{equation}
\mathcal{L}_{r}=\frac{1}{|\mathcal{T}|} \sum_{(u, v) \in \mathcal{T}}\left(r_{u, v}-\hat{r}_{u, v}\right)^{2}
\end{equation}
where where $\mathbf{W}_{1} \in \mathbb{R}^{3d \times d}, \mathbf{W}_{2} \in \mathbb{R}^{d \times 1}$ are weight parameters, $r_{u, v}$ is the ground-truth rating.

\subsection{Explanation Generation}
We use an auto-regressive style to produce words one by one, with the final output being a cohesive interpretation text. We employed a greedy decoding strategy, where the model selects the most likely word to sample at each time step. The model predicts the next hidden vector based on the currently generated one, allowing for the preservation of context throughout the generation process. 

\begin{equation}
\mathbf{e}_{t}=\operatorname{softmax}\left(\mathbf{W}^{v} \mathbf{H}_{L, t}+\mathbf{b}^{v}\right)
\end{equation}
where $\mathbf{W}^{v} \in \mathbb{R}^{|\mathcal{V}| \times d}$ and  $\mathbf{b}^{v} \in \mathbb{R}^{|\mathcal{V}|}$ are weight parameters. The vector ${e}_{t}$ represents the probability distribution over the vocabulary V.

\subsubsection{Aspect Discriminator}
To increase the probability that the selected aspects appear in explanation generation. We use the previous method \cite{ref5} and adapt it to our task. We represent $\boldsymbol{\tau}$ as the aspects that interest this user, $\boldsymbol{\tau} \in \mathbb{R}^{|\mathcal{V}|}$. If the generated word at time $t$ is an aspect, then $\tau_{a}$
is 1. Otherwise, it is 0. The loss function is as follows:
\begin{equation}
\mathcal{L}_{a}=\frac{1}{|\mathcal{T}|} \sum_{(u, v) \in \mathcal{T}} \frac{1}{\left|t_{l e n}\right|} \sum_{t=1}^{|t_{l e n}|}\left(-\tau_{a} \log e_{t, a}\right)
\end{equation}

\subsubsection{Text Generation}

We propose a mask mechanism that allows for the efficient integration of ID, aspects, and retrieved sentence information into the hidden vector of the Beginning of Sentence (BOS) position. 
At each time step, the word hidden vector is transformed into a vocabulary probability through a matrix, and the word with the highest probability is selected via the Greedy algorithm. The generation process terminates when the predicted word is the End of Sentence (EOS) marker. To ensure that the generated text adheres to a specific length, we employ a padding and truncation strategy. When the number of generated words falls short of the set length, we fill in the remaining positions with a padding token (PAD). Conversely, when the number of generated words exceeds the set length, we truncate the later words. Here we use the Negative log-likelihood loss as a generated text $\mathcal{L}_{g}$. This loss function 
 ensures the similarity between the generated words and the ground truth ones.

\begin{equation}
\mathcal{L}_{g}=\frac{1}{|\mathcal{T}|} \sum_{(u, v) \in \mathcal{T}} \frac{1}{\left|t_{len}\right|} \sum_{t=1}^{\left|t_{len}\right|}-\log e_{6+t}^{g_{ti}}
\end{equation}
where $\mathcal{T}$ denotes the training set. $g_{ti}$ denotes that use the current time $6+t$ position hidden vector to generate the $i$-th word, $i \in {1,2...,t_{len}}$. $6$ represents the initial first six positions vector information before the BOS, and $t$ represents the current moment.

\subsection{Multi-Task Learning}
We aggregate losses to form the final objective function of our multi-task learning model. The objective function of our model is defined as:
\begin{equation}
\mathcal{L}=pl\mathcal{L}_{r}+\lambda_{c}\mathcal{L}_{c}+gl\mathcal{L}_{g}+al\mathcal{L}_{a}+\lambda_{l}\|\Theta\|_{2}^{2}
\label{eq:loss_function}
\end{equation}

where $pl$, $gl$, and $al$ are the weights of the prediction and explanation, respectively. $\mathcal{L}_{g}$ represents the loss function of text generation. $\mathcal{L}_{c}$ is the loss function for context prediction. $\mathcal{L}_{a}$ is the loss function for aspect discriminator. And $\Theta$ contains all the neural parameters.

\section{Experiments}
\subsection{Datasets}
\begin{table}[t]
	\caption{Statistics of the datasets}
	\vspace{-0.2cm}
	\label{tab:datasets_result}
\resizebox{0.47\textwidth}{!}{
\begin{tabular}{l|r|r|r}
\hline Datasets & Yelp & Amazon & TripAdvisor  \\
\hline Number of users & $27{,}147$ & $157{,}212$ & $9{,}765$ \\
Number of items  & $20{,}266$ & $48{,}186$ & $6{,}280$ \\
Number of reviews & $1{,}293{,}247$ & $1{,}128{,}437$ & $320{,}023$ \\
Records per user & $47.64$ & $7.18$ & $32.77$ \\
Records per item & $63.81$ & $23.41$ & $50.96$ \\
\hline
\end{tabular}
}
\end{table}
In this part, we performed experiments on three datasets, namely Amazon (cell phones), Yelp (restaurants), and TripAdvisor (hotels) \cite{ref20}. We filtered out users with fewer than 5 comments and re-divided the dataset into three sub-datasets in the ratio of 8:1:1. The details of the datasets are shown in Table
 \ref{tab:datasets_result}. 
 We used an aspects extraction tool \cite{ref12}. The tool was used to extract the aspects in each review and correspond it to the respective review.

\begin{table}[t]
	\caption{Results of prediction}
	\vspace{-0.2cm}
	\label{tab:prediction_result}
\resizebox{0.47\textwidth}{!}{
\begin{tabular}{l|cc|cc|cc}
\hline & \multicolumn{2}{|c|}{Yelp} & \multicolumn{2}{c|}{Amazon} & \multicolumn{2}{c}{TripAdvisor } \\
\cline { 2 - 7 } & $\mathrm{R} \downarrow$ & $\mathrm{M} \downarrow$ & $\mathrm{R} \downarrow$ & $\mathrm{M} \downarrow$ & $\mathrm{R} \downarrow$ & $\mathrm{M} \downarrow$ \\
\hline PMF & $1.097$ & $0.883$ & $1.235$ & $0.913$ & $0.870$ & $0.704$ \\
SVD++ & $1.022$ & $0.793$ & $1.196$ & $0.871$ & $0.811$ & $0.623$ \\
NARRE & $1.028$ & $0.791$ & $1.176$ & $0.865$ & $0.796$ & $0.612$ \\
DAML & $1.014$ & $0.784$ & $1.173$ & $0.858$ & $0.793$ & $0.617$ \\
NRT  & $1.016$ & $0.796$ & $1.188$ & $0.853$ & $0.797$ & $0.611$ \\
CAML & $1.026$ & $0.798$ & $1.191$ & $0.878$ & $0.818$ & $0.622$ \\
PETER & $1.017$ & $0.793$ & $1.181$ & $0.863$ & $0.814$ & $0.635$ \\
ERRA & $\textbf{1.008}$ & $\textbf{0.781}$ & $\textbf{1.158}$ & $\textbf{0.832}$ & $\textbf{0.787}$ & $\textbf{0.603}$ \\
\hline
\end{tabular}
}
\end{table}

\subsection{Evaluation Metrics}
For rating prediction, in order to evaluate the recommendation performance, we employ two commonly used indicators: Root Mean Square Error (RMSE) and  Mean Absolute Error (MAE), which measure the deviation between the predicted ratings $r$ and the ground truth ratings $r^*$. For generated text, we adopted a variety of indicators that consider the quality of the generated text from different levels. 
\textbf{BLEU} \cite{ref29}, \textbf{ROUGE} \cite{ref30} and \textbf{BERTscore} \cite{ref31}  are commonly used metrics in natural language generation tasks. BLEU-N (N=1,4) mainly counts on the N-grams. 
R2-P, R2-R, R2-F, RL-P, RL-R, and RL-F denote Precision, Recall, and F1 of ROUGE-2 and ROUGE-L. BERT-S represents similarity scores using contextual embeddings to calculate. They are employed to objectively evaluate the similarity between the generated text and the targeted content.
\begin{table*}[t]
	\caption{Results of explanation}
	\vspace{-0.2cm}
	\label{tab:Explana_result}
\resizebox{16cm}{!}{
\begin{tabular}{|c|c|llll|lll|ll|}
\hline
\multirow{2}{*}{Datasets} & \multirow{2}{*}{Metrics} & \multicolumn{4}{c|}{\textbf{Baselines}}                                                                 & \multicolumn{3}{c|}{\textbf{Ours}}                                              & \multicolumn{2}{l|}{\multirow{2}{*}{Improvement}} \\ \cline{3-9}
                         &                           & \multicolumn{1}{c|}{NRT} & \multicolumn{1}{c|}{CAML} & \multicolumn{1}{c|}{ReXPlug} & \multicolumn{1}{c|}{PETER} & \multicolumn{1}{c|}{ERRA-A} & \multicolumn{1}{c|}{ERRA-R} & \multicolumn{1}{c|}{ERRA} & \multicolumn{2}{l|}{}                             \\ \hline
\multicolumn{1}{|c|}{\multirow{9}{*}{Amazon}} &
  \multicolumn{1}{l|}{BLEU1} &
  \multicolumn{1}{c|}{13.37} &
  \multicolumn{1}{c|}{11.19} &
  \multicolumn{1}{c|}{10.8} &
  \multicolumn{1}{c|}{13.78} &
  \multicolumn{1}{c|}{14.07} &
  \multicolumn{1}{c|}{13.28} &
  \multicolumn{1}{c|}{\textbf{14.38}} &
  \multicolumn{2}{c|}{4.17\%} \\ \cline{2-11} 
\multicolumn{1}{|l|}{} &
  \multicolumn{1}{l|}{BLEU4} &
  \multicolumn{1}{c|}{1.44} &
  \multicolumn{1}{c|}{1.12} &
  \multicolumn{1}{c|}{1.29} &
  \multicolumn{1}{c|}{1.68} &
  \multicolumn{1}{c|}{1.76} &
  \multicolumn{1}{c|}{1.64} &
  \multicolumn{1}{c|}{\textbf{1.88}} &
  \multicolumn{2}{c|}{10.6\%} \\ \cline{2-11} 
\multicolumn{1}{|l|}{} &
  \multicolumn{1}{l|}{R2-P} &
  \multicolumn{1}{c|}{2.06} &
  \multicolumn{1}{c|}{1.48} &
  \multicolumn{1}{c|}{2.17} &
  \multicolumn{1}{c|}{2.21} &
  \multicolumn{1}{c|}{2.67} &
  \multicolumn{1}{c|}{2.37} &
  \multicolumn{1}{c|}{\textbf{2.71}} &
  \multicolumn{2}{c|}{14.8\%} \\ \cline{2-11} 
\multicolumn{1}{|l|}{} &
  \multicolumn{1}{l|}{R2-R} &
  \multicolumn{1}{c|}{2.08} &
  \multicolumn{1}{c|}{1.23} &
  \multicolumn{1}{c|}{1.12} &
  \multicolumn{1}{c|}{2.02} &
  \multicolumn{1}{c|}{2.86} &
  \multicolumn{1}{c|}{2.33} &
  \multicolumn{1}{c|}{\textbf{2.93}} &
  \multicolumn{2}{c|}{17.6\%} \\ \cline{2-11} 
\multicolumn{1}{|l|}{} &
  \multicolumn{1}{l|}{R2-F} &
  \multicolumn{1}{c|}{1.97} &
  \multicolumn{1}{c|}{1.24} &
  \multicolumn{1}{c|}{1.22} &
  \multicolumn{1}{c|}{1.97} &
  \multicolumn{1}{c|}{2.34} &
  \multicolumn{1}{c|}{2.18} &
  \multicolumn{1}{c|}{\textbf{2.57}} &
  \multicolumn{2}{c|}{21.2\%} \\ \cline{2-11} 
\multicolumn{1}{|l|}{} &
  \multicolumn{1}{l|}{RL-P} &
  \multicolumn{1}{c|}{12.52} &
  \multicolumn{1}{c|}{9.32} &
  \multicolumn{1}{c|}{9.20} &
  \multicolumn{1}{c|}{12.62} &
  \multicolumn{1}{c|}{15.85} &
  \multicolumn{1}{c|}{13.49} &
  \multicolumn{1}{c|}{\textbf{16.13}} &
  \multicolumn{2}{c|}{19.7\%} \\ \cline{2-11} 
\multicolumn{1}{|l|}{} &
  \multicolumn{1}{l|}{RL-R} &
  \multicolumn{1}{c|}{12.20} &
  \multicolumn{1}{c|}{10.11} &
  \multicolumn{1}{c|}{10.58} &
  \multicolumn{1}{c|}{12.06} &
  \multicolumn{1}{c|}{14.11} &
  \multicolumn{1}{c|}{12.67} &
  \multicolumn{1}{c|}{\textbf{14.41}} &
  \multicolumn{2}{c|}{16.3\%} \\ \cline{2-11} 
\multicolumn{1}{|l|}{} &
  \multicolumn{1}{l|}{RL-F} &
  \multicolumn{1}{c|}{10.77} &
  \multicolumn{1}{c|}{8.11} &
  \multicolumn{1}{c|}{8.73} &
  \multicolumn{1}{c|}{11.07} &
  \multicolumn{1}{c|}{12.49} &
  \multicolumn{1}{c|}{11.97} &
  \multicolumn{1}{c|}{\textbf{13.87}} &
  \multicolumn{2}{c|}{18.1\%} \\ \cline{2-11} 
\multicolumn{1}{|l|}{} &
  \multicolumn{1}{l|}{BERT-S} &
  \multicolumn{1}{c|}{75.4} &
  \multicolumn{1}{c|}{74.9} &
  \multicolumn{1}{c|}{75.3} &
  \multicolumn{1}{c|}{76.2} &
  \multicolumn{1}{c|}{78.1} &
  \multicolumn{1}{c|}{77.3} &
  \multicolumn{1}{c|}{\textbf{79.8}} &
  \multicolumn{2}{c|}{4.5\%} \\ \hline
\multicolumn{1}{|c|}{\multirow{9}{*}{Yelp}} &
  \multicolumn{1}{l|}{BLEU1} &
  \multicolumn{1}{c|}{10.5} &
  \multicolumn{1}{c|}{9.91} &
  \multicolumn{1}{c|}{8.59} &
  \multicolumn{1}{c|}{10.29} &
  \multicolumn{1}{c|}{10.62} &
  \multicolumn{1}{c|}{10.59} &
  \multicolumn{1}{c|}{\textbf{10.71}} &
  \multicolumn{2}{c|}{3.92\%} \\ \cline{2-11} 
\multicolumn{1}{|l|}{} &
  \multicolumn{1}{l|}{BLEU4} &
  \multicolumn{1}{c|}{0.67} &
  \multicolumn{1}{c|}{0.56} &
  \multicolumn{1}{c|}{0.57} &
  \multicolumn{1}{c|}{0.69} &
  \multicolumn{1}{c|}{0.71} &
  \multicolumn{1}{c|}{0.71} &
  \multicolumn{1}{c|}{\textbf{0.73}} &
  \multicolumn{2}{c|}{5.43\%} \\ \cline{2-11} 
\multicolumn{1}{|l|}{} &
  \multicolumn{1}{l|}{R2-P} &
  \multicolumn{1}{c|}{1.95} &
  \multicolumn{1}{c|}{1.78} &
  \multicolumn{1}{c|}{1.49} &
  \multicolumn{1}{c|}{1.91} &
  \multicolumn{1}{c|}{1.95} &
  \multicolumn{1}{c|}{1.90} &
  \multicolumn{1}{c|}{\textbf{2.03}} &
  \multicolumn{2}{c|}{5.91\%} \\ \cline{2-11} 
\multicolumn{1}{|l|}{} &
  \multicolumn{1}{l|}{R2-R} &
  \multicolumn{1}{c|}{1.29} &
  \multicolumn{1}{c|}{1.05} &
  \multicolumn{1}{c|}{1.07} &
  \multicolumn{1}{c|}{1.31} &
  \multicolumn{1}{c|}{1.34} &
  \multicolumn{1}{c|}{1.29} &
  \multicolumn{1}{c|}{\textbf{1.36}} &
  \multicolumn{2}{c|}{3.6\%} \\ \cline{2-11} 
\multicolumn{1}{|l|}{} &
  \multicolumn{1}{l|}{R2-F} &
  \multicolumn{1}{c|}{1.35} &
  \multicolumn{1}{c|}{1.25} &
  \multicolumn{1}{c|}{1.11} &
  \multicolumn{1}{c|}{1.43} &
  \multicolumn{1}{c|}{1.46} &
  \multicolumn{1}{c|}{1.41} &
  \multicolumn{1}{c|}{\textbf{1.48}} &
  \multicolumn{2}{c|}{2.36\%} \\ \cline{2-11} 
\multicolumn{1}{|l|}{} &
  \multicolumn{1}{l|}{RL-P} &
  \multicolumn{1}{c|}{15.88} &
  \multicolumn{1}{c|}{14.25} &
  \multicolumn{1}{c|}{13.32} &
  \multicolumn{1}{c|}{16.07} &
  \multicolumn{1}{c|}{16.45} &
  \multicolumn{1}{c|}{15.95} &
  \multicolumn{1}{c|}{\textbf{16.60}}&
  \multicolumn{2}{c|}{3.19\%} \\ \cline{2-11} 
\multicolumn{1}{|l|}{} &
  \multicolumn{1}{l|}{RL-R} &
  \multicolumn{1}{c|}{10.72} &
  \multicolumn{1}{c|}{14.26} &
  \multicolumn{1}{c|}{9.56} &
  \multicolumn{1}{c|}{10.14} &
  \multicolumn{1}{c|}{10.83} &
  \multicolumn{1}{c|}{10.21} &
  \multicolumn{1}{c|}{\textbf{11.23}} &
  \multicolumn{2}{c|}{9.7\%} \\ \cline{2-11} 
\multicolumn{1}{|l|}{} &
  \multicolumn{1}{l|}{RL-F} &
  \multicolumn{1}{c|}{9.53} &
  \multicolumn{1}{c|}{9.16} &
  \multicolumn{1}{c|}{8.70} &
  \multicolumn{1}{c|}{10.26} &
  \multicolumn{1}{c|}{10.62} &
  \multicolumn{1}{c|}{10.14} &
  \multicolumn{1}{c|}{\textbf{10.82}} &
  \multicolumn{2}{c|}{5.1\%} \\ \cline{2-11} 
\multicolumn{1}{|l|}{} &
  \multicolumn{1}{l|}{BERT-S} &
  \multicolumn{1}{c|}{83.6} &
  \multicolumn{1}{c|}{83.2} &
  \multicolumn{1}{c|}{82.2} &
  \multicolumn{1}{c|}{83.3} &
  \multicolumn{1}{c|}{84.7} &
  \multicolumn{1}{c|}{83.1} &
  \multicolumn{1}{c|}{\textbf{85.2}} &
  \multicolumn{2}{c|}{2.2\%} \\ \hline
\multicolumn{1}{|c|}{\multirow{9}{*}{TripAdvisor}} &
  \multicolumn{1}{l|}{BLEU1} &
  \multicolumn{1}{c|}{15.78} &
  \multicolumn{1}{c|}{14.43} &
  \multicolumn{1}{c|}{12.64} &
  \multicolumn{1}{c|}{15.33} &
  \multicolumn{1}{c|}{15.93} &
  \multicolumn{1}{c|}{15.43} &
  \multicolumn{1}{c|}{\textbf{16.13}} &
  \multicolumn{2}{c|}{5.9\%} \\ \cline{2-11} 
\multicolumn{1}{|l|}{} &
  \multicolumn{1}{l|}{BLEU4} &
  \multicolumn{1}{c|}{0.85} &
  \multicolumn{1}{c|}{0.86} &
  \multicolumn{1}{c|}{0.71} &
  \multicolumn{1}{c|}{0.89} &
  \multicolumn{1}{c|}{1.02} &
  \multicolumn{1}{c|}{0.95} &
  \multicolumn{1}{c|}{\textbf{1.06}} &
  \multicolumn{2}{c|}{15.8\%} \\ \cline{2-11} 
\multicolumn{1}{|l|}{} &
  \multicolumn{1}{l|}{R2-P} &
  \multicolumn{1}{c|}{1.98} &
  \multicolumn{1}{c|}{1.49} &
  \multicolumn{1}{c|}{1.61} &
  \multicolumn{1}{c|}{1.92} &
  \multicolumn{1}{c|}{2.03} &
  \multicolumn{1}{c|}{1.97} &
  \multicolumn{1}{c|}{\textbf{2.09}} &
  \multicolumn{2}{c|}{8.1\%} \\ \cline{2-11} 
\multicolumn{1}{|l|}{} &
  \multicolumn{1}{l|}{R2-R} &
  \multicolumn{1}{c|}{1.92} &
  \multicolumn{1}{c|}{1.91} &
  \multicolumn{1}{c|}{1.49} &
  \multicolumn{1}{c|}{2.01} &
  \multicolumn{1}{c|}{2.1} &
  \multicolumn{1}{c|}{1.98} &
  \multicolumn{1}{c|}{\textbf{2.15}} &
  \multicolumn{2}{c|}{9.7\%} \\ \cline{2-11} 
\multicolumn{1}{|l|}{} &
  \multicolumn{1}{l|}{R2-F} &
  \multicolumn{1}{c|}{1.9} &
  \multicolumn{1}{c|}{1.92} &
  \multicolumn{1}{c|}{1.61} &
  \multicolumn{1}{c|}{1.94} &
  \multicolumn{1}{c|}{2.02} &
  \multicolumn{1}{c|}{1.99} &
  \multicolumn{1}{c|}{\textbf{2.05}} &
  \multicolumn{2}{c|}{5.3\%} \\ \cline{2-11} 
\multicolumn{1}{|l|}{} &
  \multicolumn{1}{l|}{RL-P} &
  \multicolumn{1}{c|}{14.85} &
  \multicolumn{1}{c|}{13.36} &
  \multicolumn{1}{c|}{11.38} &
  \multicolumn{1}{c|}{13.54} &
  \multicolumn{1}{c|}{15.3} &
  \multicolumn{1}{c|}{14.84} &
  \multicolumn{1}{c|}{\textbf{15.40}} &
  \multicolumn{2}{c|}{8.6\%} \\ \cline{2-11} 
\multicolumn{1}{|l|}{} &
  \multicolumn{1}{l|}{RL-R} &
  \multicolumn{1}{c|}{14.03} &
  \multicolumn{1}{c|}{12.38} &
  \multicolumn{1}{c|}{10.22} &
  \multicolumn{1}{c|}{14.75} &
  \multicolumn{1}{c|}{14.93} &
  \multicolumn{1}{c|}{14.77} &
  \multicolumn{1}{c|}{\textbf{15.02}} &
  \multicolumn{2}{c|}{1.81\%} \\ \cline{2-11} 
\multicolumn{1}{|l|}{} &
  \multicolumn{1}{l|}{RL-F} &
  \multicolumn{1}{c|}{12.25} &
  \multicolumn{1}{c|}{12.39} &
  \multicolumn{1}{c|}{9.97} &
  \multicolumn{1}{c|}{12.61} &
  \multicolumn{1}{c|}{13.08} &
  \multicolumn{1}{c|}{12.79} &
  \multicolumn{1}{c|}{\textbf{13.17}} &
  \multicolumn{2}{c|}{4.50\%} \\ \cline{2-11} 
\multicolumn{1}{|l|}{} &
  \multicolumn{1}{l|}{BERT-S} &
  \multicolumn{1}{c|}{82.7} &
  \multicolumn{1}{c|}{84.8} &
  \multicolumn{1}{c|}{83.2} &
  \multicolumn{1}{c|}{86.4} &
  \multicolumn{1}{c|}{87.6} &
  \multicolumn{1}{c|}{86.9} &
  \multicolumn{1}{c|}{\textbf{88.1}} &
  \multicolumn{2}{c|}{1.96\%} \\ \hline
 % &
 %   &
 %   &
 %   &
 %   &
 %   &
 %   &
 %   &
 %   &
  % \multicolumn{2}{l}{}
\end{tabular}
}
\end{table*}

\subsection{Baseline Methods}
\subsubsection{Prediction}
The performance in terms of accuracy of rating prediction is compared with two types of methods: Machine Learning and Deep Learning: 

\begin{itemize}[leftmargin=*]
    \item Deep learning models: \textbf{NARRE} \cite{ref4} is a popular type of neural network for text-based tasks. \textbf{PETER} \cite{ref9} and \textbf{NRT} \cite{ref23} are deep learning models that use review text for prediction and explanation at the same time.
    \item Factorization methods: \textbf{PMF} \cite{ref27}  is a matrix factorization method that uses latent factors to represent users and \textbf{SVD++} \cite{ref28} leverages a user's interacted items to enhance the latent factors.
\end{itemize}

\begin{table}[!t]
\renewcommand\arraystretch{1.1}
\centering

\label{tab:verify concepts}

\end{table}

\subsubsection{Explainability}
To evaluate the performance of explainability, we compare against three explanation methods, namely CAML \cite{ref5} and ReXPlug \cite{ref22} and NRT and PETER. 

\begin{itemize}[leftmargin=*]
\item \textbf{ReXPlug} uses GPT-2 to generate texts and is capable of rating prediction.
\item \textbf{CAML} uses users' historical reviews to represent users and uses co-attention mechanisms to pick the most relevant reviews and concepts and combine these concepts to generate text. 
    \item \textbf{NRT} is an advanced deep learning method for explanation tasks. As a generative method, NRT mainly generates explanations based on predicted ratings and the distribution of words in tips.
    \item \textbf{PETER} is a powerful model improved by a transformer. This model effectively integrates the ID in the transformer and combines this ID information as the starting vector to generate text.
\end{itemize}
\vspace{0cm}

\subsection{Reproducibility}
We conducted experiments by randomly splitting the dataset into a training set (80\%), validation set (10\%), and test set (10\%).
The baselines were tuned by following the corresponding papers to ensure the best results. The embedded vector dimension was 384 and the value yielded superior performance after conducting a grid search within the range of [128, 256, 384, 512, 768, 1024]. The maximum length of the generated sentence was set to 15-20.
The weight of the rating prediction ($pl$) was set to 0.2, and the weight of the $\lambda_{c}$ and $al$ was set to either 0.8 or 0.05. For the explanation task, the parameter $gl$ was adjusted to 1.0 and was initialized using the Xavier method \cite{ref35}. The models were optimized using the Adam optimizer with a learning rate of $10^{-1}$ and L2 regularization of $10^{-4}$. When the model reached the minimum loss in a certain epoch, the learning rate would be changed at that time and multiplied by 0.25. When the total loss of continuous three epochs has not decreased, the training would stop. More implementation details are on github\footnote{https://github.com/Complex-data/ERRA}.

\subsection{Explainability Study}
\textbf{Explainability results:} Table \ref{tab:Explana_result} shows that our proposed ERRA method consistently outperforms the baselines in terms of BLEU and ROUGE on different datasets. For instance, take BLEU as an example, our method demonstrates the largest improvement on the TripAdvisor dataset. It is likely due to the smaller size of the dataset and the relatively short length of the reviews, which allows for additional information from the retrieved sentences and aspects to supplement the generated sentences, leading to an enhancement in their richness and accuracy. In contrast, the increase in BLEU on the Yelp dataset is relatively small. It is due to the large size of the Yelp dataset, which allows the model to be trained on a vast amount of data. The GPT \cite{ref24} series also prove this case, large amounts of data can train the model well, resulting in our retrieval not having as obvious an improvement compared to other datasets.

Similarly, when compared with NRT and PETER, our model consistently outperforms them in all metrics. Whether it is in terms of the fluency of the sentence, the richness of the text, or the consistency with the real label, our model has achieved excellent results.

\noindent\textbf{Case study:} 
We take three cases generated from three datasets by NRT, PETER, and ERRA method as examples. Table \ref{tab:explanation_case}
shows that the ERRA model can predict keywords, which are both closer to the original text and match the consumers' opinions, generating better explanations compared to the baseline. 
While the baseline model always generates statements and explanations that are not specific and detailed enough, our model can generate personalized, targeted text, such as \emph{the battery doesn't last long} in Case 2 and \emph{excellent! The food here is very delicious!} in Case 3. This either is the same as or similar to the ground truth.

\noindent\textbf{Human evaluation:}
We also evaluate the model's usefulness in generated sentences via the fluency evaluation experiment, which is done by human judgment. We randomly selected 1000 samples and invited 10 annotators to assign scores. Five points mean very satisfied, and 1 point means very bad. Table \ref{tab:fluency_evaluation} reports the human evaluation results. Kappa \cite{ref37} is an indicator for measuring classification accuracy. Results demonstrate that our model outperforms the other three methods on fluency and Kappa metrics. 
\begin{figure}[tbp]
	\small
	\centering
        \subfloat[]{\includegraphics[width=0.25\textwidth]{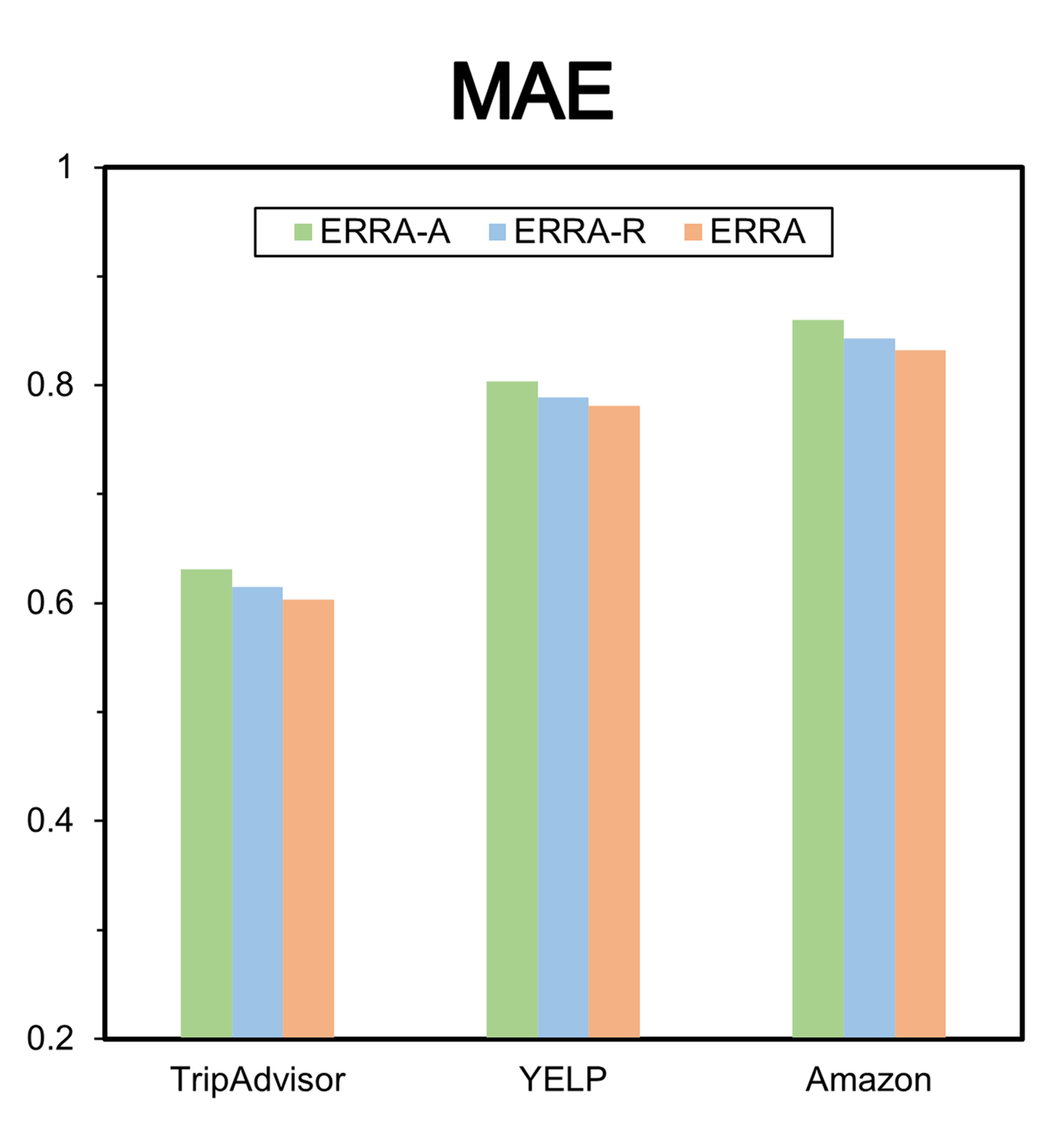}}
        \subfloat[]{\includegraphics[width=0.25\textwidth]{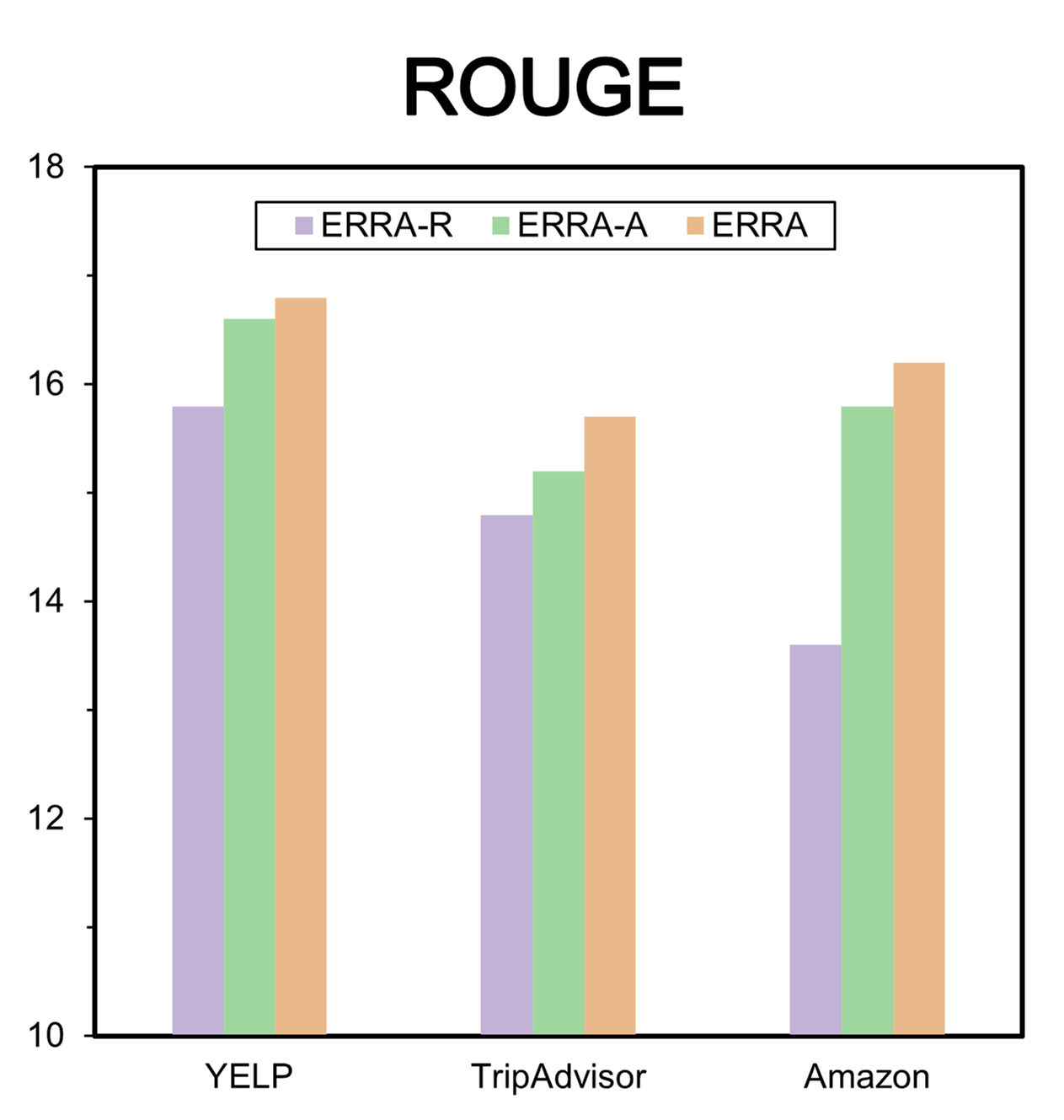}}
\caption{Ablation analysis of prediction and explanation tasks}
\label{fig:Ablation}
\end{figure}
\subsection{Accuracy of Prediction}

The evaluation result of prediction accuracy is shown in Table \ref{tab:prediction_result}. As we can see, it shows that our method consistently outperforms baseline methods including  PMF, NRT, and PETER in RMSE and MSE for all datasets. We mainly compare the performance of our model with the PETER model, which is a state-of-the-art method. Our model demonstrates a significant improvement over the baseline methods on the TripAdvisor dataset. We attribute this improvement to the way we model users. By taking aspects into consideration, our model is capable of accurately modeling users. And this in turn can generate more accurate predictions. As shown in Table \ref{tab:prediction_result}, ERRA's predictive indicator is the best result on each dataset.

\subsection{Ablation Analysis}
In order to investigate the contribution of individual modules in our proposed model, we performed ablation studies by removing the retrieval enhancement and aspect module denoted as "ERRA-R" and "ERRA-A",  From Figure \ref{fig:Ablation}(a), we can see that the retrieval module plays a crucial role in enhancing the performance of the explanation generation task. Specifically, for the Amazon and TripAdvisor datasets, the difference between "ERRA-R" and ERRA is the largest for explanation generation, while showing mediocrity in the  prediction task.

Additionally, we also evaluated the impact of the aspect enhancement module on performance. Figure \ref{fig:Ablation}(b) shows that removing this module leads to a visible decline in both the prediction and explanation tasks. It is due to different users having different attention points, and the aspects can more accurately represent the user's preference, thus making the prediction more accurate and the generated text more personalized.

\begin{table}[t]
%\LARGE
\centering
\caption{Explanations generated by ERRA and Baseline.}
\vspace{-0.2cm}
\label{tab:explanation_case}
\resizebox{0.45\textwidth}{!}{
\begin{tabular}{|p{0.08\textwidth}|p{0.37\textwidth}|}
%\begin{tabular}{|l|l|}
\hline
Case 1 - Truth  & The environment of this hotel is comfortable and the transportation is very convenient and the sound insulation effect is great. \textcolor{blue}{Aspects}:\textcolor{blue}{(environment, comfortable)} \textcolor{blue}{(hotel, insulation)}
\\ \hline
NRT  & The environment of this hotel is best! \\ \hline
PETER & The hotel service is pretty good! looks very nice!
% I wouldn't come again. 
\\ \hline

ERRA & The room \textbf{environment} is pretty \textbf{comfortable}! The  \textbf{traffic} here is very  \textbf{convenient}.
% I wouldn't come again. 
\\ \hline
\hline
Case 2 - Truth & 
The screen of this phone is too small and his battery drains fast so I can't stand it.
  \textcolor{blue}{Aspects}:\textcolor{blue}{(screen, too small)} \textcolor{blue}{(battery, fast)}
% (Aspects:\textbf{too small},\textbf{fast})
\\ \hline
NRT & The phone is bad. 
\\ \hline

PETER & The phone is bad,  It works poorly and I don't like it.
\\ \hline
ERRA & I really \textbf{hate} this phone, the \textbf{battery} \textbf{doesn't last long}, the \textbf{screen} is \textbf{faulty}.
\\ \hline\hline
Case 3 - Truth  & Delicious! The customer service is pretty good and the open all the way to 3 am in the morning. The prime foods are excellent! 
  \textcolor{blue}{Aspects}:\textcolor{blue}{(service, good)} \textcolor{blue}{(foods, excellent)}
\\ \hline
NRT  & The service is pretty good.  \\ \hline
PETER & he tastes delicious! The service is pretty good!
\\ \hline

ERRA & \textbf{excellent}! The \textbf{service} here is \textbf{pretty good}. The food  here is \textbf{very delicious}! There are many unique foods in it and open \textbf{till dawn}.
\\ \hline

\end{tabular}}
\end{table}

\begin{table}[tp]
\small
\centering
\caption{Results of the fluency evaluation.}
\vspace{-0.2cm}
\label{tab:fluency_evaluation}
\begin{tabular}{ccccc}
\hline
Measures & NRT  & CAML & ReXPlug & ERRA \\ \hline
Fluency  & 2.73 & 2.92 & 3.11   & \textbf{3.45} \\
Kappa & (0.67) & (0.63) & (0.74)   & (\textbf{0.79})  \\ \hline
\end{tabular}
\end{table}

\section{Conclusion}
In this paper, we propose a novel model, called  ERRA, that integrates personalized aspect selection and retrieval enhancement for prediction and explanation tasks. To address the issue of incorrect embedding induced by data sparsity, we incorporate personalized aspect information and rich review knowledge corpus into our model. Experimental results demonstrate that our approach is highly effective compared with state-of-the-art baselines on both the accuracy of recommendations and the quality of corresponding explanations.
%\bibliography{ACL}
\bibliographystyle{unsrt}

\appendix

\end{document}